\documentclass[aps,prl,twocolumn,showpacs,superscriptaddress]{revtex4}
\usepackage{graphics}
\usepackage{amsmath}
\usepackage{epsfig}

\begin{document}
\title{Quantum Monte Carlo Simulation of the High-Pressure Molecular-Atomic Crossover in Fluid Hydrogen}

\author{Kris~T.~Delaney}
\affiliation{Department~of~Physics, University~of~Illinois~at~Urbana-Champaign,
Urbana, IL 61801, USA}
\author{Carlo~Pierleoni}
\affiliation{INFM-SOFT and Dipartimento~di~Fisica, Universit\`a~dell'Aquila, I-67010 L'Aquila, Italy}
\author{D.~M.~Ceperley}
\affiliation{Department~of~Physics, University~of~Illinois~at~Urbana-Champaign,
Urbana, IL 61801, USA}

\date{\today}

\begin{abstract}
A first-order liquid-liquid phase transition in high-pressure hydrogen between 
molecular and atomic fluid
phases has been predicted in computer simulations using {\it ab
initio} molecular dynamics approaches. However, experiments
indicate that molecular dissociation may occur through a continuous crossover rather
than a first-order transition. Here we study the nature of molecular
dissociation in fluid hydrogen using an alternative simulation
technique in which electronic correlation is computed within
quantum Monte Carlo, the so-called Coupled Electron Ion Monte
Carlo (CEIMC) method. We find no evidence for a first-order
liquid-liquid phase transition.
\end{abstract}


\maketitle


The behavior of hydrogen under a range of thermodynamic conditions
is an important problem in theoretical physics, as well as in
planetary science and high-pressure physics. Hydrogen exhibits a
rich variety of properties including several molecular and
non-molecular phases,\cite{HReview} a fluid metal-insulator
transition,\cite{shockwave} and possible fluid\cite{Hquantumfluid}
and superconducting\cite{Hsuperconduct} phases at low temperature.
There exists much uncertainty about the equilibrium properties of
high-pressure phases, including the structure and relative
stabilities of a number of solid phases,\cite{HSolidPhases,HSolidPhases2} the
specific shape of the molecular-solid melting curve
and the reason for the density
maximum,\cite{Hquantumfluid} and the conditions of molecular
dissociation in the fluid.

Shock-wave experiments \cite{shockwave} have hinted that the
metalization of hydrogen in the liquid state occurs in a
continuous manner at temperatures above $3000\mathrm K$. However,
recent computer simulations\cite{ScandoloPNAS03,Hquantumfluid}
using Car-Parrinello molecular dynamics (CPMD) with density
functional theory within the local density approximation (DFT-LDA)
and generalized gradient approximation (DFT-GGA) predict that the
molecular dissociation in the fluid happens through a first-order phase
transition, with discontinuous metalization, at temperatures lower
than $3000\mathrm{K}$. These results could be reconciled by having
the critical point for the transition at $T < 3000\mathrm{K}$.
In addition, other simulations of hydrogen fluid phases, which
do not directly address the nature of the molecular dissociation, have been
reported \cite{HFluidSimsNOPPT}.

In this Letter, we present the results of a study of this
transition employing a new quantum Monte Carlo (QMC) method,
coupled electron-ion Monte Carlo (CEIMC).\cite{ceimc1,ceimc2} The
first QMC approaches for studying high pressure hydrogen were limited to
studying solid phases at $T=0\mathrm{K}$: variational Monte Carlo (VMC) and diffusion
quantum Monte Carlo (DMC)\cite{T0HHdiss,RPAJastrow}. Finite-temperature
restricted path integral Monte Carlo (RPIMC) was used
to investigate the possibility of a hydrogen liquid-liquid phase transition
\cite{pimcppt}, finding a continuous molecular dissociation at
$T \simeq 10000\mathrm{K}$ in agreement with experiments.\cite{shockwave}
However, RPIMC becomes inefficient and inaccurate at lower
temperatures.

The CEIMC approach, in contrast with earlier QMC methods, involves
simultaneous evolution of separate but coupled Monte Carlo
simulations for the electronic and ionic subsystems, which may
then be treated on different levels of approximation. A large gain
in computational efficiency can be made for lower temperatures by
employing the Born-Oppenheimer approximation (BOA) and requiring
that the electronic subsystem remain in its ground state
($T=0\mathrm{K}$). The ionic system is treated at finite
temperature either as a set of classical particles, or quantum
mechanically using the imaginary-time path integral formalism.
This decoupling is good provided the thermal excitation of
electrons is a small effect, a fact that holds if $T\ll T_F$,
where $T_F$ is the Fermi temperature. For the densities of
interest, those at which molecules dissociate in the fluid, $T_F >
200,000\mathrm{K}$. The quality of the BOA applied to high-pressure
hydrogen fluid has been tested, and found to be good, by comparing
CEIMC to accurate RPIMC calculations, which contain no such
approximation, at temperatures of 5,000 and
10,000K.\cite{CarloH2Melt} We therefore anticipate that the BOA
will be even more accurate at lower temperatures.

For classical nuclei in the canonical ensemble, one can generate
the configuration-space probability distribution by proposing a
move from configuration $S$\cite{defineS} to configuration
$S^\prime$ with uniform probability and using the Metropolis
acceptance probability
\begin{equation}
A=\mathrm{min}\left[1,\mathrm{exp}\left(-\beta\Delta\left(S,S^\prime\right)\right)\right],
\end{equation}
where $\beta$ is the inverse temperature,
$\left(k_BT\right)^{-1}$, and $\Delta\left(S,S^\prime\right)$ the
difference in BO energies of nuclear configurations $S$ and
$S^\prime$, computed with a $T=0\mathrm{K}$ QMC method. The bias resulting
from the statistical noise of $\Delta$ is reduced with correlated
sampling\cite{ceimc1,ceimc2}, and eliminated by using the penalty
method \cite{penalty}.

For electronic sampling, we use either VMC or reptation quantum
Monte Carlo (RQMC). RQMC \cite{rqmc} is a projector method that
does not suffer from the mixed-estimator bias of DMC and allows
efficient calculations of energy differences. To handle electron
antisymmetry, we use the fixed-phase method \cite{fixedphase}
which allows the modulus of the many-electron wavefunction to be
fully optimized while the phase is held fixed. This gives an
estimate of the ground state energy much lower than that of the
variational estimate (by typically 2500K/atom) but above that of
the exact ground state energy. Analysis of related electron
systems suggest that the error of the fixed-phase energy will be
between 10\% and 30\% of the VMC error,\cite{kwon} i.\,e.,
between 250K/atom and 750K/atom. We expect relative errors,
between different proton configurations, to be even
less.\cite{C184} We also employ twist-averaged boundary conditions
(TABC)\cite{TABC} in the electronic calculation to greatly
reduce finite-size effects, a problem that has caused substantial
errors in earlier simulations of fluid hydrogen.\cite{Hohl}
Well-converged energies and correlation functions are achieved
using, depending on the density, between 108 and 500 twist angles.

\begin{figure}
  \resizebox{\columnwidth}{!}{\rotatebox{-90}{\includegraphics{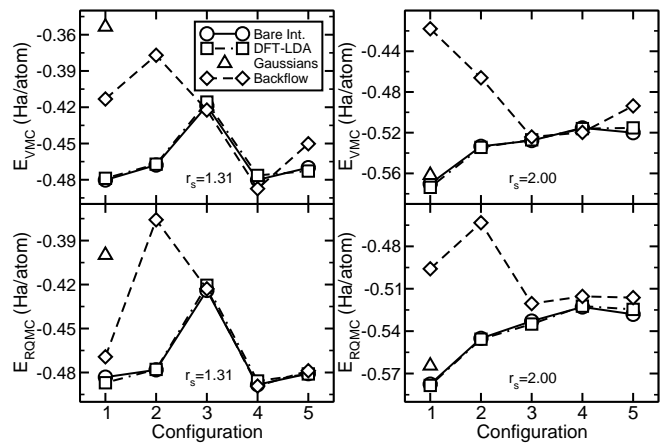}}}
  \caption{VMC and RQMC total energies for five different crystal configurations at two different
    densities (left graphs $V=9.42 \mathrm{a.u./atom}$ or $r_s=1.31$; right graphs $V=33.51 \mathrm{a.u./atom}$ or $r_s=2.0$)
    using a number of different Slater-Jastrow wavefunctions. Configuration 1 is molecular,
    2--5 are non-molecular. Owing to the variational principle, a lower energy implies a
    better wavefunction. DFT-LDA and bare electron-proton bands (see text) provide the
    best and most transferable orbitals for the Slater determinant. Gaussian for configuration
    1 refers to localized molecular orbitals.}
  \label{fig:wfntest}
\end{figure}

In the CEIMC approach, one is free to choose any trial
wavefunction for the electronic ground-state. An important
consideration is the computer time required per step, since the wavefunction
must be calculated for many nuclear configurations. When studying dense
liquid hydrogen, we require a general wavefunction that is equally
accurate for both molecular and non-molecular configurations.
Therefore, we employ a fast band-structure calculation with an
effective one-electron potential designed to produce accurate
single-particle orbitals for a Slater-Jastrow type wavefunction.
Within the fixed-phase approach, it is only the orbitals that
affect the systematic bias. One such set of orbitals would be
those computed within Kohn-Sham theory. However, full
self-consistency for each proton displacement would be too
expensive for generating a large number of nuclear
configuration-space samples. Consequently, we use a bare
electron-proton potential for the effective single-particle
potential. Further screening and correlation effects are
introduced with a Slater-Jastrow wavefunction; the Jastrow factor
is from the RPA pseudopotential, including the one-body (electron-proton)
term.\cite{RPAJastrow} Such an approach is surprisingly good, as
demonstrated by Figure \ref{fig:wfntest} which shows VMC and RQMC total
energies for five different frozen nuclear configurations. This
trial function is comparable in quality to one using Kohn-Sham LDA
orbitals in the Slater determinant for all configurations and densities tested,
while being faster to generate. It is also more transferable than both
localized molecular orbitals\cite{ceimc1} (which cannot be used for
non-molecular systems) and analytic backflow\cite{backflow} (which is highly accurate
only for high-density metallic systems).

\begin{figure}
  \resizebox{\columnwidth}{!}{{\rotatebox{-90}{\includegraphics{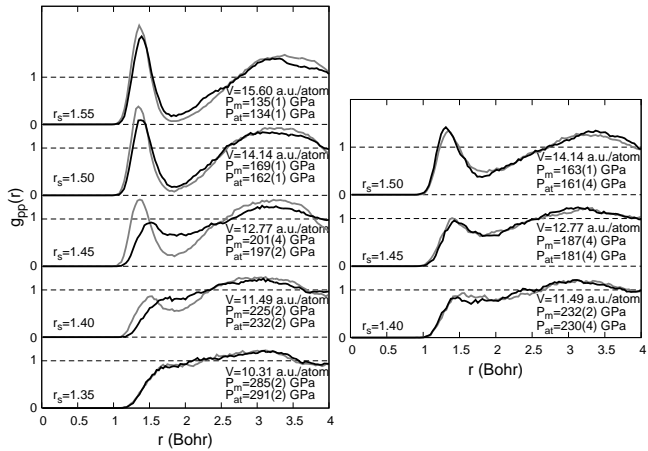}}}}
  \caption{Proton-proton correlation functions for CEIMC simulations in the canonical ensemble with 32 atoms at $T=2000\mathrm{K}$. The BO energies are computed with VMC (left pane) or RQMC (right pane) and the wavefunction is a Slater-Jastrow type with bare electron-proton bands (see text). Grey lines are simulations initially prepared with a molecular fluid and black with an atomic fluid. Hysteresis is evident. $P_\mathrm{at}$ and $P_\mathrm{m}$ are the computed pressures for simulations prepared with an atomic and a molecular fluid respectively. All correlation functions are plotted to the same scale.}
  \label{fig:t2kgrs}
\end{figure}

We investigate the atomic-molecular transition in liquid hydrogen
at fixed density ($1.35 \leq r_s \leq 1.55$ where $4 \pi
r_s^3/3=1/n_e$) and temperature ($T=2000\mathrm{K}$ and $1500\mathrm{K}$).
The pressure, estimated using the virial theorem, an approach that is accurate
with RQMC due to the lack of a mixed-estimator bias, lies between
135GPa and 290GPa for this range of densities.
Using a classical Monte Carlo simulation with
an empirical potential,\cite{ceimc2} the system is prepared
either in a purely molecular or a purely atomic fluid initial
configuration. During the subsequent simulation the system evolves
to its equilibrium state, subject to overcoming any free-energy
barriers during the simulation. A typical simulation has 14,000
ionic moves with a step size of 0.006--0.016Bohr. We collect
statistics along the sequence of ionic states, particularly the
proton-proton correlation functions which give insight into the
state of the liquid through a distinctive peak at
$r_{pp} \sim 1.4 \mathrm{Bohr}$ when molecules are present. Figure
\ref{fig:t2kgrs} shows a set of proton-proton correlation
functions for simulations at several densities prepared with
either a molecular or atomic fluid as the initial state.

\begin{figure}
  \resizebox{0.85\columnwidth}{!}{\rotatebox{-90}{\includegraphics{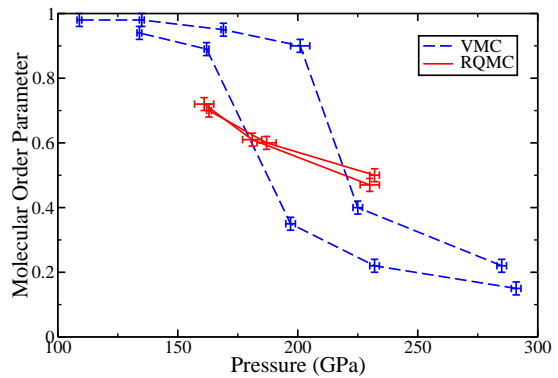}}}
  \caption{Molecular order parameter ($\lambda$ in Eq.\ \ref{eq:mop}) evaluated for simulation results from Fig.\ \ref{fig:t2kgrs}. Two curves are obtained for each method from different initial conditions (see text). VMC (dashed lines) simulations indicate an irreversible, weakly first-order phase transition for molecular dissociation in the fluid with increasing/decreasing pressure. This picture is compatible with conclusions from CPMD DFT-LDA. Accurate RQMC (solid lines) simulations find a reversible crossover.}
  \label{fig:orderparam}
\end{figure}

For a quantitative analysis, a method for estimating the average
number of molecules from the pair-correlation function at each phase point
is required. A rough estimate would involve integrating the pair-correlation
function up to the first minimum, but such an
approach does not take into account the ``baseline'' caused by nearby molecules and unbound atoms; see Fig. \ref{fig:t2kgrs}. To remove the baseline, we fit each of
the pair-correlation functions to the functional form:
\begin{equation}
  g\left(r\right) = \lambda g_\mathrm{m}\left(r;\left\{\alpha\right\}\right) + \left(1-\lambda\right)g_\mathrm{at}\left(r;\left\{\gamma\right\}\right),
  \label{eq:mop}
\end{equation}
where $\left\{\alpha,\gamma\right\}$ are fitting parameters and
$g_\mathrm{m}$ is a Gaussian centered on 1.3--1.5Bohr.
$g_\mathrm{at}$ is a fit to the next-nearest neighbor
distribution.
We find that the quality of the fit is insensitive to the choice of
$g_\mathrm{at}$; the molecular fraction varies by no more than
$3\%$ for different choices. The molecular order parameter,
$\lambda$, the fraction of protons that are bound into
a molecule, is plotted against pressure in Fig.
\ref{fig:orderparam} for $T=2000\mathrm{K}$.

We find that the CEIMC simulations using VMC energy differences
(left pane Fig.~\ref{fig:t2kgrs}, and Fig.~\ref{fig:orderparam}) yield an
irreversible transition with hysteresis. This implies a
free-energy barrier that is difficult to overcome during the
simulations, hence a metastable state is obtained. The large width of the
hysteresis loop is indicative of a transition that is weakly
first-order. We therefore expect that $T=2000\mathrm{K}$ is close to the VMC
critical point.

In contrast, the much more accurate RQMC simulations (right pane
Fig.~\ref{fig:t2kgrs}, and Fig.~\ref{fig:orderparam}) yield a reversible and
continuous molecular dissociation upon increasing pressure. Both
molecular and atomic states are mechanically unstable in the range
of densities simulated, and the system quickly evolves, with no
discernible free-energy barrier, to a stable part-molecular,
part-atomic state. This behavior implies the lack of an underlying
first-order transition, a behavior which would not change as the
thermodynamic limit is approached. The likely reasons for the
change in character of the crossover is that the BO energy
surface of the assumed Slater-Jastrow trial function is inaccurate
under conditions at which molecules are short-lived transient entities.
As discussed above, these deficiencies are corrected by the
imaginary-time projection of RQMC, if they are due to errors in the
modulus of the trial wavefunction and not its phase.

\begin{figure}
  \resizebox{0.65\columnwidth}{!}{{\includegraphics{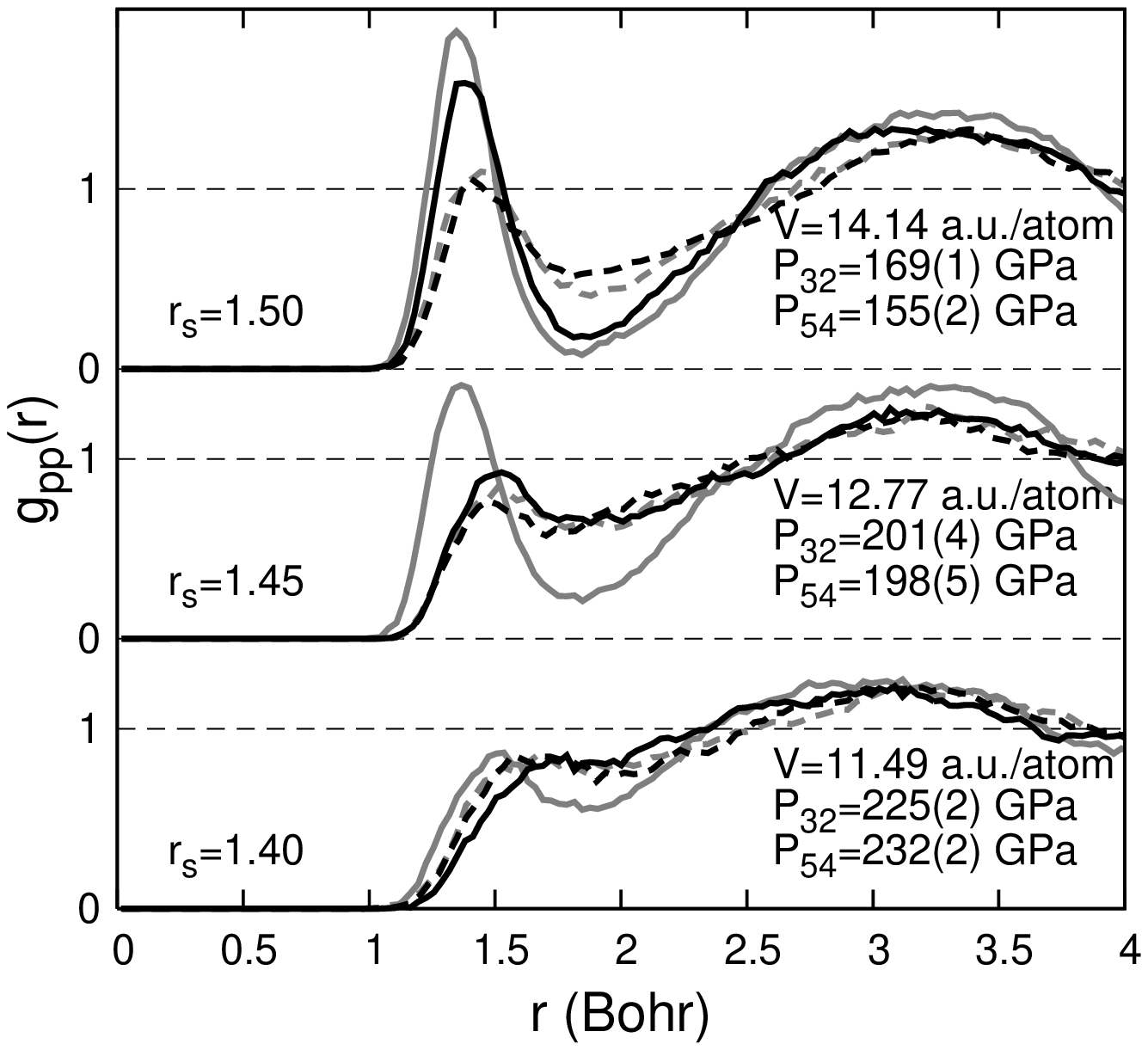}}}
  \resizebox{0.65\columnwidth}{!}{{\includegraphics{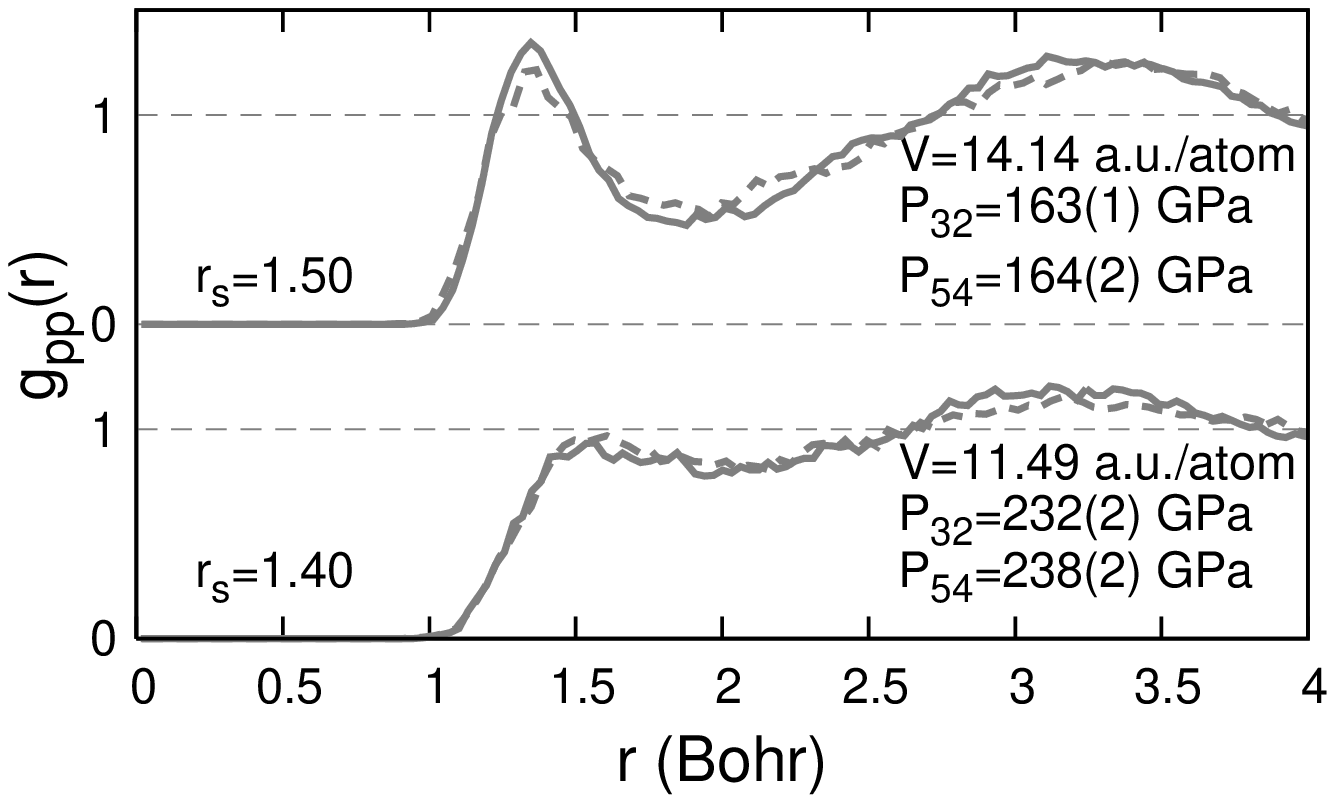}}}
  \caption{Demonstration of finite-size errors in VMC pair-correlation functions (upper pane). Solid lines are 32-atom simulations, dashed are 54 atoms. Grey lines are simulations started from a molecular fluid and black are from an atomic fluid. Quoted pressures are for simulations started from a molecular fluid. RQMC (lower pane) simulations show small finite-size errors.}
  \label{fig:fse}
\end{figure}

We have assessed the finite-size errors in the cell volume by
repeating a selection of simulations with 54 atoms. The
proton-proton correlation functions in Figure \ref{fig:fse} show
the comparison between 32- and 54-atom simulations for VMC and
RQMC. For VMC finite-size errors are clearly large, consistent
with the presence of a free-energy barrier, while for RQMC the errors are
small, indicative of a continuous crossover between molecular and
atomic states. 

\begin{figure}
  \resizebox{\columnwidth}{!}{\rotatebox{-90}{\includegraphics{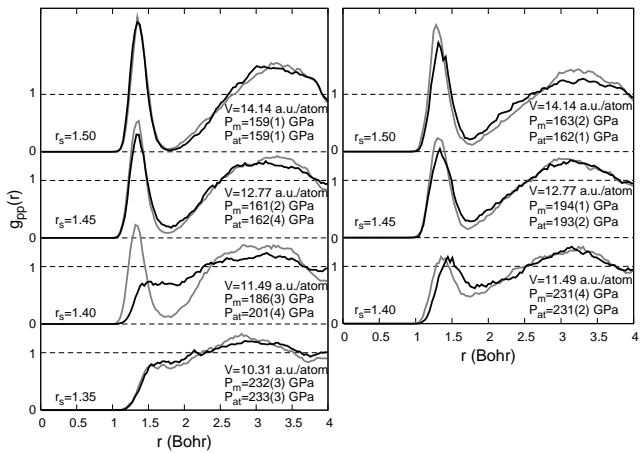}}}
  \caption{Proton-proton correlation functions for simulations at $T=1500\mathrm{K}$. Black lines are simulations prepared with an atomic fluid, grey are prepared with a molecular fluid. The simulations demonstrate an irreversible transition with hysteresis for VMC. Left pane: VMC; Right pane: RQMC.}
  \label{fig:t1500grs}
\end{figure}

We have tested the nature of molecular dissociation at a lower
temperature: $T=1500\mathrm{K}$. Figure \ref{fig:t1500grs} shows
pair-correlation functions. Hysteresis is again evident for VMC
simulations, although over a shorter pressure-range, indicating a
larger free-energy barrier. Again, RQMC simulations do not show
a first-order transition, mirroring the behavior at 
$T=2000\mathrm{K}$.


In conclusion, this study demonstrates that the CEIMC approach
with a fast and accurate trial wavefunction can be used to
determine the nature of the pressure-driven molecular to atomic
crossover in hydrogen for temperatures on the order of 1000K. We
find that the improvements in interatomic interactions obtained
when using RQMC to simulate the electronic subsystem lead to no
first-order phase transition, a result of fundamentally different 
nature to that predicted when using VMC. Our most accurate simulations, 
those employing RQMC, provide no evidence for a first-order atomic-molecular
transition in the liquid phase at either $T=1500\mathrm{K}$ 
or $T=2000\mathrm{K}$, showing instead a continuous crossover;
this is in contrast to the first-order transition found
with CPMD. The remaining uncontrolled approximations in our
simulations are the finite-size error (demonstrated to be small
for RQMC), the fixed-phase approximation in RQMC, the adiabatic
approximation for decoupling nuclear and electronic time-scales
(tested in ref. \cite{CarloH2Melt}) and the use of classical
statistics for simulating nuclei. Path integral calculations are
in progress for studying the effect of nuclear quantum statistics
on dense liquid hydrogen. We anticipate that nuclear quantum
effects will destabilize molecules at a lower pressure leaving the
crossover character unchanged.

\begin{acknowledgments}
This material is based upon work supported by the NSF awards DMR 04-04853
and DMR 03-25939 ITR, and by MIUR-COFIN2003. Calculations were
performed at the NCSA at the University of Illinois at
Urbana-Champaign and at CINECA (Italy). We thank S. Chiesa and D.
Quigley for fruitful discussions.
\end{acknowledgments}

\end{document}